# The BUCEA Speaker Diarization System for the VoxCeleb Speaker Recognition Challenge 2022

*Ruohua Zhou, Yuxuan Du, Chenlei Hu*

School of Electrical and Information Engineering, Beijing University of Civil Engineering and Architecture, China

{2108110020016}@bucea.edu.cn

## Abstract

This paper describes the BUCEA speaker diarization system for the 2022 VoxCeleb Speaker Recognition Challenge. Voxsrc-22 provides the development set and test set of VoxConverse, and we mainly use the test set of VoxConverse for parameter adjustment. Our system consists of several modules, including speech activity detection (VAD), speaker embedding extractor, clustering methods, overlapping speech detection (OSD), and result fusion. Without considering overlap, the Dover-LAP (short for Diarization Output Voting Error Reduction) method was applied to system fusion, and overlapping speech detection and processing were finally carried out. Our best system achieves a diarization error rate (DER) of 5.48% and a Jaccard error rate (JER) of 32.1% on the VoxSRC 2022 evaluation set respectively.

## 1. Introduction

The task of speaker diarization is to break up multi-speaker audio into homogenous single speaker segments, effectively solving 'who spoke when' [1]. And The goal of Voxsrc-22 challenge is to probe how well current methods can segment and recognize speakers from speech obtained 'in the wild'. The dataset is obtained from YouTube videos, consisting of multi-speaker audio from both professionally edited videos as well as more casual conversational multi-speaker audio in which background noise, laughter, and other artefacts are observed in a range of recording environments.

VoxConverse [2] is an audio-visual diarization dataset consisting of over 50 hours of multispeaker clips of human speech, extracted from YouTube videos. Voxsrc-22 provides VoxConverse as the validation set. VoxConverse includes the development set and the test set. Through our test, the speaker confusion part in the diarization error rate of the test set is worse than the verification set, which is more suitable for our parameter adjustment.

In this report, we will introduce our diarization models respectively and present experimental results of our submission to the speaker diarization track (track 4). We describe our system details in Section 2. We present experimental results and have a brief analysis in Section 3, and summarize our conclusions in Section 4.

## 2. System Description

### 2.1 Overview

Our proposed speaker diarization system is shown in Figure 1. The VAD model first processes the input audio to obtain valid speech segments. Then all segments are uniformly divided at different time scales. For each time scale, the ResNet model is used to extract the speaker embeddings. Next, different methods are used to cluster the speaker embeddings, and diarization outputs are fused by DOVER-Lap. Finally, the audio overlap is processed. The details of each module are explained in the following subsections.

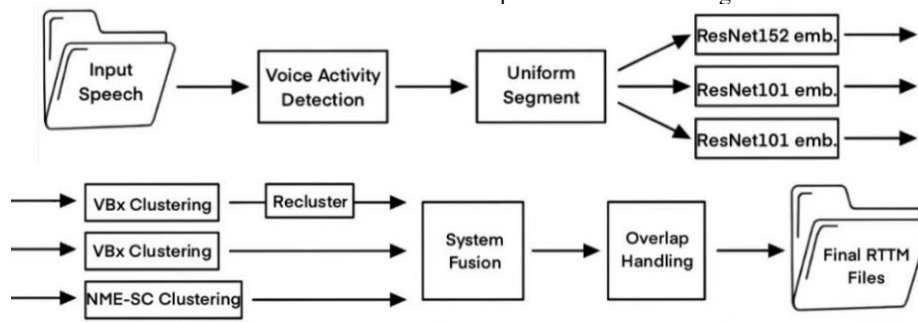

Figure 1: System overview

### 2.2 Voice activity detection

VAD is the first and most critical step of the whole diarization system. Due to the complex noise environment of the final evaluation set from VoxSRC, it is easy to hurt the final clustering results. We first utilized the VAD toolkit from Kaldi to perform energy-based VAD, which didn't have a good effect. Later, we used the toolkit Pyannote 2.0 [3]. We fine-tuned the parameters and found that the result is better when the speech interval is less than 0.7 seconds. We also tried to apply speech enhancement to reduce noise to improve the performance of VAD, but the results were not satisfactory. Unfortunately, we don't have more time to compare the performance of VAD utilizing the Resnet34 model [4].

Specifically, Pyannote benefits from having a post-processing step that fills gaps shorter than a given threshold and removes active regions shorter than another given threshold.

Table1 shows the VAD results for dev1 and dev2 separately.

Table 1: The false alarm (FA), miss detection (MISS) and total error of the VAD model

| Test set | MISS | FALSE | ERROR |
|---|---|---|---|
| DEV1 | 2.6 | 0.3 | 2.9 |
| DEV2 | 2.9 | 0.8 | 3.7 |

### 2.3 Segmentation

In our system, we utilized the uniform segmentation method. The choice of frame length and shift is mainly based on the parameters of the BUT team [5]. We tried to fine-tune the parameters and got better results in the development set. However, the result is not reliable in the evaluation set, so we still apply the original parameters. In the case of fixed window size, the longer the frame length, the more information about the speaker's identity can be obtained, and the shorter the frame shift, the more accurate the speaker's representation is. Due to the frame shift is the smallest unit to determine the speaker's identity, a smaller frame shift help pinpoint speaker change points.

### 2.4 Speaker Embedding

A speaker embedding extractor is employed to convert acoustic features into fixed-dimensional feature vectors. We mainly apply ResNet101 and ResNet152 [6] as the speaker embedding extractors which input is a 64-dimensional log Mel filter bank feature with a 25ms frame length and 10ms frame shift. The audio is cut into 1.44 second segments in our system, and the frame shift is 0.18/0.24 seconds. The 16 kHz x-vector extractor is trained using data from VoxCeleb1, VoxCeleb2 [7], and CN-CELEB. Among them, VoxCeleb has a duration of 1323 hours with 1211 speakers, VoxCeleb2 has a duration of 2291 hours with 5994 speakers, and CN-CELEB has a duration of 264 hours with 973 speakers. Additionally, the data is augmented from the MUSAN [8] and RIR [9] corpora.

### 2.5 Clustering

*2.5.1 Initial Clustering*

Moreover, we evaluate different clustering methods, including agglomerative hierarchical clustering (AHC) and normalized maximum eigenspectral clustering (NME-SC) [10]. In the subsequent reclustering process, the combination of the AHC with VBx [11] can make the single system achieve the best performance. Through experimental analysis, the speaker embeddings extracted by the Resnet101 extractor are reduced by Linear Discriminant Analysis (LDA) from 256-dimensional speaker embeddings to 128-dimensions without changing the LDA dimensionality of the Resnet152 extractor to get better results. The cosine similarity matrix is then used as the input to the AHC model, and the threshold is set to -0.015.

*2.5.2 Re-Clustering*

The VBx model is detailed in [11] to improve the performance of initial clustering. The results of the initial clustering are used for VBx model initialization. Variable Bayes HMM at the level of the x-vector BHMM is used to cluster the x-vector. The HMM states represent speakers, the transitions between states represent speaker turns, and a PLDA [12] model derives the state distribution pretrained on the labeled x-vector. The hyperparameters mentioned in the VBx model were tuned on the Voxconverse dev, where $Fa=0.3$, $Fb=16$, $loopP=0.9$. In the recluster period, we apply the method that proposed by the BUT team. We concatenate the initial rttm results to extract new embeddings. Each speaker's global x-vectors are clustered with AHC to join speakers if necessary. Then reclustering the results using different clusters separately, and the performance is shown in table 2.

### 2.6 DOVER-Lap based system fusion

DOVER-Lap [13] is a system fusion algorithm for combining the outputs of multiple journaling systems. The algorithm applies a global mapping strategy based on the cost tensor. System fusion corrects erroneous speaker labels for a single system and detects segments with overlapping speakers by voting on segment labels generated by different diary systems. We tried to fuse different clustering methods in the development set and found that the results of VBx are more suitable for fusion with the results of spectral clustering. The results after fusion are shown in table 2.

Table 2: Performance on development set and eval set (only system 10 handling overlaps), with diarization error rate (DER) and Jaccard error rate (JER) shown in percentage (%).

| System | Embedding | Cluster | Recluster | Dev2 DER | Dev2 JER | Eval DER | Eval JER |
|---|---|---|---|---|---|---|---|
| 1 | | Baseline | | - | - | 19.60 | 41.43 |
| 2 | ResNet101 | AHC | No | 7.34 | 30.50 | - | - |
| 3 | ResNet101 | VBx | No | 5.62 | 32.46 | 6.00 | 30.03 |
| 4 | ResNet101 | NME-SC | No | 7.56 | 37.07 | - | - |
| 5 | ResNet152 | AHC | Yes | 7.13 | **25.25** | - | - |
| 6 | ResNet152 | VBx | Yes | **5.56** | 28.55 | - | - |
| 7 | ResNet152 | NME-SC | Yes | 6.42 | 37.22 | - | - |
| 8 | | Fusion 2+3+4 | | 5.60 | 29.80 | - | - |
| 9 | | Fusion 5+6+7 | | 5.54 | **27.56** | - | - |
| 10 | | Fusion 3+4+6 | | **5.32** | 30.94 | **5.48** | 32.1 |

**2.7 Overlap speech detection**

Overlapping speech detection (OSD) is the task of estimating the area where two or more speakers speak simultaneously. It accounts for a large proportion of the various error rates of the diarization system, so the processing of overlapping speech is essential. Pyannote, as a diarization open-source toolkit, includes an overlapping speech detection model. The model in pyannote is trained on a composite training set of DIHARD3 [14], AMI [15], and dev1. Here we only discuss the case where the overlapping segment includes two speakers. Assuming that only two people are speaking simultaneously in the overlapping segment, the speaker should be the closest in the time series before and after the overlapping segment. Therefore, we assign the overlap of speakers to the nearest speaker [16]. Also, unlike other teams, we fused the system first and then dealt with the overlap, due to we found that this performed better in the dev set than fusing with overlapping labels.

## 3. Result and Analysis

According to our results, a single system achieves the best performance on the evaluation set when using Resnet152 to extract speaker embeddings and clustering using the VBx method. Although changing the frameshift on the dev set gives better results, this doesn't seem to apply to the evaluation set. In the spectral clustering method, we try to use a known number of people in the development set to cluster, and manually adjust the threshold, but it does not work well. Likewise, although the speaker embedding codes extracted by the Res152 network for spectral clustering and reclustering showed the best results of this method, the results were not significantly improved after fusion.

Furthermore, although the system fusion results on DER have improved, the reasons for the decline in JER results still require further analysis.

## 4. Conclusion

This report presents the diarization system we submitted for the diarization task (Track 4) of the 2022 VoxCeleb Speaker Recognition Challenge. We applied a fusion of different clustering methods and the Dover Lap system to further improve the performance of our original module-based system. Based on the evaluation set in the final submission, we achieved a diarization error rate (DER) of 5.48% and a Jaccard error rate (JER) of 32.1%.

Diarization is still a challenging task at present, and requires close cooperation between various modules. Our future work is to research short-duration speech speaker recognition.